\documentclass[pra,aps,showpacs, twocolumn]{revtex4-1}
\usepackage{mathtext}
\usepackage[T2A]{fontenc}
\usepackage[cp1251]{inputenc}
\usepackage{epsf}
\usepackage{psfrag}
\usepackage{graphicx}
\usepackage{amssymb,amsmath}
\usepackage[usenames]{color}
\usepackage{colortbl}

\begin{document}

\title{Motional effects in dynamics of fluorescence of cold atomic ensembles excited by resonance pulse radiation}
\author{A. S. Kuraptsev}
\email[]{aleksej-kurapcev@yandex.ru}
\affiliation{\small Peter the Great St. Petersburg Polytechnic University, 195251, St. Petersburg, Russia}
\author{I. M. Sokolov}
\email[]{sokolov_im@spbstu.ru}
\affiliation{\small Peter the Great St. Petersburg Polytechnic University, 195251, St. Petersburg, Russia}

\date{\today}

\sloppy



\begin{abstract}
We report the investigation of the influence of atomic motion on the fluorescence dynamics of dilute atomic ensemble driven by resonant pulse radiation. We show that even for sub-Doppler temperatures, the motion of atoms can significantly affect the nature of both superradiation and subradiation. We also demonstrate that, in the case of an ensemble of moving scatterers, it is possible to observe the nonmonotonic time dependence of the fluorescence rate. This leads to the fact that, in certain time intervals, increasing in temperature causes not an decrease but  increase of the fluorescence intensity in the cone of coherent scattering. We have analyzed the role of the frequency diffusion of secondary radiation as a result of multiple light scattering in an optically dense medium. It is shown that spectrum broadening is the main factor which determines radiation trapping upon resonant excitation. At later time, after the trapping stage, the dynamics is dominated by close pairs of atoms (dimers). The dynamics of the excited states of these dimers has been studied in detail. It is shown that the change in the lifetime of the given adiabatic term of the diatomic quasi-molecule induced by the change in the interatomic distance as well as possible non-adiabatic transitions between sub- and superradiant states caused by atomic motion can lead not to the anticipated weakening of subradiation effect but to its enhancement.

\end{abstract}
\maketitle

\section{Introduction}
Atomic ensembles cooled to sub-Doppler temperatures in special traps are currently of great interest both because of
 the number of their unique physical properties, and because of the wide range of their possible practical applications in problems of quantum metrology, frequency standardization, quantum information applications \cite{1,2,3}.

Almost all proposed schemes for the use of cold atomic ensembles as well as most diagnostic methods  are based on interaction of these  ensembles with electromagnetic radiation. This interaction has a number of features associated with collective polyatomic effects due to
low speed of the atoms.  These effects are due, first, to the large resonant cross sections for light scattering by each separate atoms and, consequently, to the large optical depth of the ensembles even at low atomic densities. The second reason is random spatial disorder, in which the formation of atomic clusters, or quasi-molecules, consisting of several atoms, randomly located at distances of the order of the resonance radiation wavelength from each other, is possible. Dipole-dipole interatomic interaction causes the formation of collective sub- and superradiant states, which can essentially affect the optical properties of cold gases.

The main approach to the description of collective effects is now the so-called method of coupled oscillators. To date, several variants of this method have been developed \cite{F45, L51, Javanainen:1999, RMO00, PRMOT00, 34a, Svidzinsky:2010, 38a,KSH11,Bienaime:2013,SS14, BGAK14, Guerin:2016a, KS_PRA_2016,Guerin:2017a,Kuraptsev:2017,Skipetrov_2019}. The main difficulty in using this method is accounting for the motion of atoms in real physical systems. Therefore, in the overwhelming majority of works, the approximation of fixed scatterers is used. The displacement of atoms is taken into account by averaging the observables over a random spatial distribution of atoms. In papers \cite{JRSY14,JRJBPSB16,BZBZBSNKYLRY16} an attempt was made to refine the immobile atom approximation. In the refined model, the Doppler shift was modeled by introducing a random shift in the frequencies of atomic transitions, different for different atoms.

The effect of continuous displacement of atoms in dilute media was considered in the framework of the scalar approximation in \cite{10}. A more detailed experimental and theoretical analysis is given in \cite{11}. The main result of this work was the assertion that subradiative states are sufficiently resistant to thermal decoherence at the temperatures of magneto-optical trap (MOT). Similar stability is predicted up to temperatures on the order of mikelvin. We came to a different conclusion in our group when considering dense atomic ensembles with a strong dipole-dipole interatomic interaction \cite{12}. For clouds in which the average interatomic distance is comparable with the wavelength of resonant radiation, we observed the destruction of subradiative states even at temperatures several times lower than the typical MOT temperatures.

The essential influence of motion on another collective effect, on the effect of single-photon superradiance, was discovered in the framework of the study of the flash effect in the works \cite{14,15}. Here, in particular, it was shown that the subradiation rate in the direction of the exciting pulse increases upon heating. For a flat layer of atoms for an infinitesimal time interval after the end of the excitation pulse, it was even possible to obtain analytical expressions confirming this growth. At the same time, theoretical studies of superradiance outside the cone of coherent forward scattering, carried out in \cite{16}, lead to opposite conclusions. Heating manifests itself in a negative way, weakening the superradiance in these directions.

Thus, the available data indicate the complex nature of the influence of atomic motion on collective optical effects. This influence depends both on the nature of the effect and on the conditions of observation. The main purpose of this work is to study unexplored case of dilute atomic ensembles cooled to sub-Doppler temperatures excited by resonance pulse radiation.  Within the framework of a unified approach based on the coupled oscillator method accounting for continuous displacement, we will consider atomic fluorescence in a wide time interval and study the features of both superradiance and subradiation at different temperatures. We will show that even when the characteristic Doppler frequency shifts are smaller than the natural width of atomic transitions, motion can significantly affect the fluorescence dynamics.

\section{Basic assumptions and approach}
In our theoretical description of time-dependent fluorescence we use the coupled dipoles  model, which is traditional for this class of the problems \cite{F45, L51, Javanainen:1999, RMO00, PRMOT00, 34a, Svidzinsky:2010, 38a,KSH11,Bienaime:2013,SS14, BGAK14, Guerin:2016a, KS_PRA_2016,Guerin:2017a,Kuraptsev:2017,Skipetrov_2019}.

We consider a disordered atomic ensemble of $N$  identical two-level atoms. All atoms have a ground state $|g\rangle$ with the total angular momentum $J_g = 0$, an excited state $|e\rangle$ with $J_e = 1$, a transition frequency $\omega_0$, and a natural lifetime of all excited Zeeman sublevels ($ m  = -1,0,1 $) is $\tau_0 = 1/\gamma$.

Our specific calculations is based on approach developed earlier in \cite{KSH11, 12}. In accordance with this approach we study the properties of a closed system consisting of all atoms and an electromagnetic field, including a vacuum reservoir. We seek the wave function $ \psi $ of this system as an expansion over the eigenfunctions $ {\psi_l} $ of the Hamiltonian of noninteracting atoms and light $ \psi = \sum_l \beta_l \psi_l $. Assuming that the exciting radiation is weak, which is typical in experiments \cite{6,7}, we take into account only states with no more than one photon in the field. In such a case for the amplitudes $ \beta_e $ of one-fold excited atomic states $ \psi_e = | g \cdots e \cdots g \rangle $ we have the following differential equations
\begin{equation}
\frac{\partial \beta_e}{\partial t} = \left( i\delta-\frac{\gamma}{2} \right)\beta_e -\frac{i\Omega_{e}}{2} + \frac{i\gamma}{2} \sum_{e' \neq e} V_{ee'}\beta_{e'}.
\label{e1}
\end{equation}
Here, the index $ e $ shows both the number of the atom which  is excited in the state $ \psi_e = | g \cdots e \cdots g \rangle $, and specific Zeeman sublevel populated in this state; $\Omega_e$ is the Rabi frequency of the external laser field in the point where atom $e$ locates, $\delta$ is the detuning of the field from resonance atomic frequency.

The last term in Eq. (\ref{e1}) corresponds to dipole-dipole interatomic interaction and is responsible for  collective effects in the considered atomic ensemble. The matrix $V_{ee'}$ is
\begin{eqnarray}
V_{ee'}& =&
-\frac{2}{\gamma} \sum\limits_{\mu, \nu}
\mathbf{d}_{e g}^{\mu} \mathbf{d}_{g e'}^{\nu}
\frac{e^{i k_0 r_{ij}}}{\hbar r_{ij}^3}
\nonumber
\\
&\times& \left\{
\vphantom{\frac{r_{ij}^{\mu} r_{ij}^{\nu}}{r_{ij}^2}}
 \delta_{\mu \nu}
\left[ 1 - i k_0 r_{ij} - (k_0 r_{ij})^2 \right]
\right.
 \\
&-&\left. \frac{\mathbf{r}_{ij}^{\mu} \mathbf{r}_{ij}^{\nu}}{r_{ij}^2}
\left[3 - 3 i k_0 r_{ij} - (k_0 r_{ij})^2 \right]
\right\}.
\nonumber
\end{eqnarray}
Here we assume that in the states $e$ and $e'$ atoms $i$ and $j$ are excited; $\mathbf{d}_{e g}$ is the matrix element of the dipole moment operator for the transition ${g} \to {e}$, $\mathbf{r}_{ij} =\mathbf{r}_i - \mathbf{r}_j$, $r_{ij} = |\mathbf{r}_i - \mathbf{r}_j|$ and
$k_0=\omega_0/c$ is the wavenumber associated to the transition, with $c$ the vacuum speed of light. The indexes $\mu$ and $\nu$ denote projections of vectors on the axes of the reference frame.

From the values of $\beta_{e}(t)$ computed on the basis of Eq. (\ref{e1})  we can find  the amplitudes of all other states which determine the wave function $\psi$ (for more detail see \cite{KSH11}), which in its turn gives us information about the properties of the secondary radiation as well as about the the properties of atomic ensemble. In particular, the intensity $I_\alpha (\mathbf{\Omega},t )$ of the light  polarization component $\alpha$ that the atoms scatter in a unit solid angle around direction of the wave vector $\mathbf{k}$  determined by radius-vector $\mathbf{r}$ ($\mathbf{\Omega}={\theta,\varphi}$) reads
\begin{eqnarray} I_\alpha (\mathbf{\Omega},t )&=&\frac{c}{4\pi}  \left\langle \psi
\right\vert E^{(-)}_\alpha(\mathbf{r}) E^{(+)}_\alpha(\mathbf{r}) \left\vert \psi \right\rangle r^2 \label{e2}
\\
&=&\frac{c}{4\pi} \left|  k_{0}^{2} \sum\limits_{e}\left(\mathbf{u}_\alpha^*\mathbf{d}_{ge}\right)  \beta_{e}(t) \exp \left(
-i\mathbf{k}\mathbf{r}_{i})\right)  \right| ^{2}. \nonumber
\end{eqnarray}
Here $E^{(\pm)}_\alpha(\mathbf{r})$ are the positive and negative frequency parts of the electric field operator; $\mathbf{u}_\alpha$ is the unit polarization vector of the scattered light.

In this paper, while analyzing the role of atomic motion, we will not conduct a detailed study of the angular distribution of fluorescence. The main attention will be attend to the study of the influence of temperature on the dynamics of the total radiation of the ensemble. This value can be obtained by integrating the expression (\ref{e2}) over the total solid angle and summing the contributions of the various polarization components. It can also be calculated on the basis of the law of energy conservation, taking into account that the total radiation energy is equal to the decrease in the excitation energy of the atomic system, i.e. it can be determined from the rate of decrease in the total population $P_{ex}(t)$ of the excited states of all atoms. The latter can be found  as follows
\begin{equation}
P_{ex}(t)=\sum\limits_{e}\left\vert \beta_{e}(t)\right\vert ^{2}.
\label{e3}
\end{equation}

In the next section, based on the relations (\ref{e1})-(\ref{e3}), we will calculate the rate of decay of the total radiation intensity of an ensemble of moving atoms at different temperatures. We will look for a non-stationary solution to Eq. (1), taking into account the displacement of atoms with time explicitly. We will consider the temperature ranges typical for MOT and higher, at which the momenta of atoms are much greater than the momenta of a photon. For this reason, and also taking into account the weakness of the excitation, we will not take into account the recoil effects and will describe the motion as a classical uniform and rectilinear motion $ \mathbf {r} _i = \mathbf {r} _{i0} + \mathbf {v}_i (t-t_0) $. In order not to take into account departure of atoms from the considered volume and associated change in their densities, we will assume that the volume of the cloud is surrounded by imaginary surfaces which scatters the atoms elastically without modification of its internal states. For simplify we consider an ensemble having the shape of a cube with edge equal to $L$.

The distribution of atoms at  initial time $t=t_0$ is considered random, but spatially homogeneous on average. The atomic medium is optically dense but dilute. The density of atoms $n$ in all calculations will be the same $nk_0^{-3}=0.005$. The average distance between atoms in this case exceeds the wavelength of quasi-resonant radiation.

The velocities of the atoms at $t=t_0$ are also considered as random variables. All their projections $ v_\mu $  are assumed to be distributed according to the gaussian law.
\begin{equation}\label{3}
f(v_\mu)=1/\sqrt{2\pi v_0^2}\exp(-v_\mu^2/2v_0^2).
\end{equation}
The dispersion of the velocities $ v_0 $ and the wave number $ k_0 $ determine the Doppler broadening of the line $ \Delta_D = 2 \sqrt {2 \ln2} k_0v_0 $. All fluorescence parameters calculated in this paper will be obtained by averaging over random variables $\mathbf {r} _{i0}$ and $ \mathbf {v}_i$.

The radiation pulse that excites fluorescence will be considered rectangular, its carrier frequency is resonant to the transition in a free atom $\delta=0$. For definiteness, we choose it to be right-circularly polarized.

\section{Results}

As the main quantity characterizing the dynamic of the fluorescence, we will use the current (instantaneous) radiation delay time $\tau(t)=1/\Gamma(t)$, where $\Gamma (t)=d\ln(I(t))/dt$, and $I(t)$ is the total intensity of the secondary radiation of the atomic ensemble. The dependence $\tau(t)$ after the end of the excitation pulse with the duration $\gamma T=50$ at different temperatures of the ensemble containing $N=625$ atoms is shown in Fig. 1.
\begin{figure}
     \centering
     \includegraphics[width=0.5\textwidth]{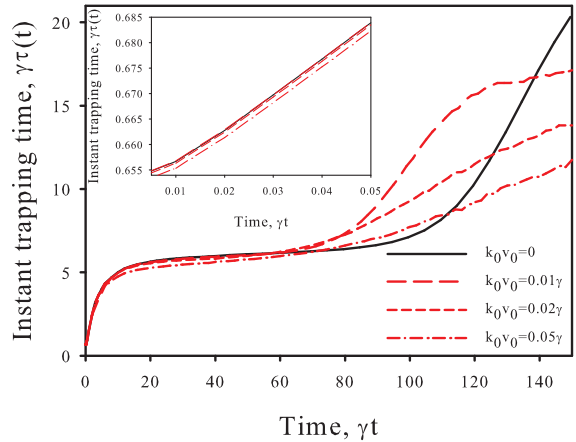}
\caption{Dynamics of instantaneous fluorescence delay time at various $k_0v_0$ (temperatures). The number of atoms is $N=625$. Excitation pulse duration is $\gamma T=50$.}
\end{figure}
As for immobile atoms, \cite{22,23} several characteristic stages of fluorescence can be distinguished. First, at times $t<1/\gamma$ after the end of the excitation pulse, the superradiance effect is observed. The dependence $\tau(t)$ at this stage is shown on an enlarged scale in the inset to Fig. 1. The decay rate $\Gamma(t)$ here  is greater than the natural width $\gamma$, and $\gamma\tau(t)<1$.

Then comes the stage of radiation trapping, which is due to the diffusion of photons in an optically dense medium. It can be divided into two parts. Initially, the decay rate decreases, and the trapping time increases. Here, radiation diffusion is described by multimode dynamics. Further, the diffusion regime becomes single-mode, when the afterglow decay is described with good accuracy by a single-exponential law. This regime corresponds to rectilinear, almost horizontal segments on the $\tau(t)$ curves.

Finally, after the one-exponential phase, a noticeable increase in the trapping time $\tau(t)$ and a decrease in the decay rate are observed. Here we are dealing with the radiation of clusters randomly formed in the considered disordered ensemble. These clusters have long-lived states that are responsible for the "classical" $\;$ subradiation process predicted by Dicke \cite{Dicke}.

Next, we consider in more detail those features of the fluorescence dynamics that result from taking into account the motion of atoms at each of these main stages.

\subsection{Influence of motion on the nature of single-photon superradiance}
As already mentioned, the effect of single-photon superradiance has been studied in sufficient detail. The dynamics of fluorescence in the cone of coherent forward scattering has been studied especially detailed. In particular, in the experiment \cite{14} it was found that the rate of superradiance in this direction increases with heating. This unexpected effect has been explained as the result of the dephasing effect from the motion of the atoms \cite{15}. In the model of a flat layer, infinite in the transverse direction, for the rate of the initial stage corresponding to the time $t=0^+$ immediately after the abrupt switching off of the excitation, an analytical expression is obtained, which in the case of resonant excitation has the form

\begin{equation}\label{4}
\Gamma(0^+)=\frac{b_0 \gamma}{2 (1-exp(-b(v_0)/2))}.
\end{equation}
Here $b(v_0)$ and $b_0$  are the optical thickness of the medium at a given temperature, and at a temperature tending to zero respectively.
\begin{figure}
     \centering
     \includegraphics[width=0.5\textwidth]{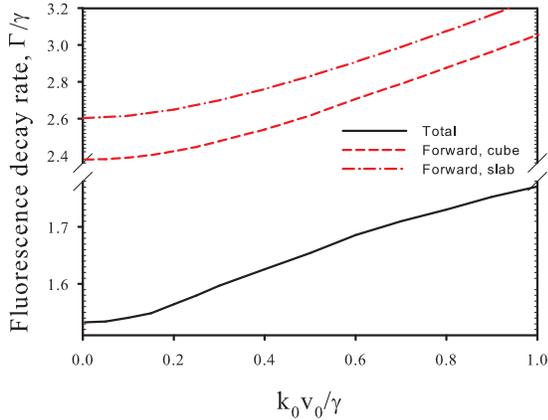}
    \caption{Dependence of the average fluorescence rate $\Gamma $ in the time interval $\Delta t=0-0.01\gamma $ on temperature ($k_0v_0$). Number of atoms is $N=625$, density is $nk_0^{-3}=0.005$. The black solid line corresponds to the total radiation intensity in all directions and polarizations, the red dotted line corresponds to forward radiation, in the direction of the wave vector of the exciting light. The red dash-dotted line is calculated based on the Eq. (\ref{4}). }
\end{figure}

Fig. 2 shows the results of our numerical calculation for a cloud of finite size. Here, the dependence on temperature (more precisely, on the value of $k_0v_0$) of the average fluorescence rate $\Gamma$ is shown. Averaging was carried out for a finite time interval of duration $0.01\, \gamma$ after the end of the excitation pulse. The black solid line was obtained for the total radiation intensity over all directions and polarizations. On a different scale this curve demonstrates what corresponds to the region of small times in Fig. 1.

\begin{figure}
     \centering
     \includegraphics[width=0.5\textwidth]{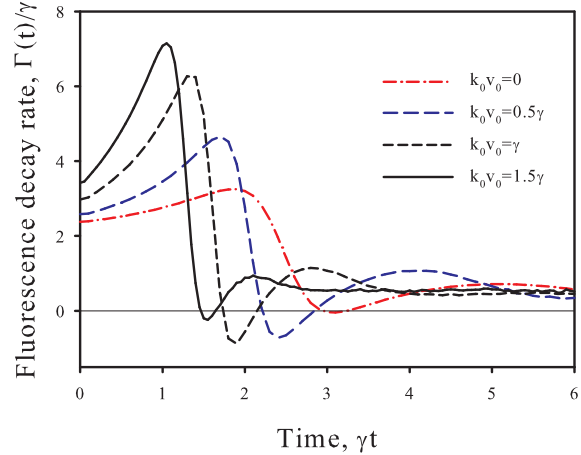}
    \caption{Time dependence of the fluorescence rate in the forward direction $\Gamma (t)$ at different temperatures ($k_0v_0$). }
\end{figure}

The temperature dependence of $\Gamma$ for radiation in the forward scattering lobe is shown in Fig. 2 by the red dotted line. In the considered range of parameters, this fluorescence rate is more than one and a half times higher than the average in all directions. The red dash-dotted line is calculated based on the Eq. (\ref{4}). It can be seen that the flat layer model reproduces the qualitative temperature dependence of $\Gamma (t)$ quite well, but leads to noticeable quantitative differences  for the scatterer ensemble with finite transverse dimensions.

Numerical analysis of the forward fluorescence process revealed an important feature of the time dependence of $\Gamma (t)$ on time intervals of the order of the natural lifetime of an atomic excited state. Accounting for motion leads to a qualitative change in the dynamics of fluorescence. Its speed is not maximum at $t=0^+$. After the excitation is turned off, it changes nonmonotonically (see Fig. 3). It can even change the sign. I.e. at certain time intervals, the intensity of secondary radiation in the coherent forward lobe does not decrease, but increases. The velocity $\Gamma$ reaches its maximal negative values at $k_0v_0\sim \gamma$.  To the best of our knowledge, this effect has not been observed for immobile atoms.

In our opinion the oscillation in the afterglow of the atomic ensemble connectes with quantum beating and is caused
by interference of light scattering through different collective states \cite{24,12}.

\subsection{Influence of motion on diffusion trapping}
As follows from Fig. 3, for the ensembles under study, the transient processes end at the time $\gamma t\sim 5-6$ after the superradiance stage. Then the trapping stage begins. Here the trapping time changes slightly with temperature, which, in our opinion, is precisely what was observed in the  experiment \cite{11}.

This result seems quite natural, since, as is known, the trapping time $\tau_d$, given by the horizontal segment in Fig. 1  is proportional to the square of the optical thickness $b$. For immobile atoms and clouds of large optical thickness $b_0\gg 1$ this time is well described by the following simple relation
\begin{equation}\label{5}
\tau_d=\frac{3b_0^2}{\alpha\pi^2}\tau_{0},
\end{equation}
where the parameter $\alpha$ depends on the shape of the cloud. For cubic volume $\alpha=3$.

The decrease in $\tau_d$ is associated with a change in the mean free path due to the Doppler effect. The role of this effect is relatively small at sub-Doppler temperatures. However, numerical calculations show that this decrease turns out to be more significant than the relation (\ref{5}) predicts if $b_0$ is replaced by $b(v)$ for moving atoms.

The solid and dotted black lines in Fig. 4 show the calculated dependences of $\tau_d$ on the atomic velocity for two sizes of the atomic ensemble $k_0L=50$ and $k_0L=60$. The density of atoms in both cases is the same and equal to $nk_0^{-3}=0.005$. The red dashed-dotted lines show how the time $\tau_d$ would change if it was calculated using the formula (\ref{5}) taking into account the dependence of the optical thickness on temperature. For the convenience of comparison, the results calculated by the Eq. (\ref{5}) were renormalized so that they coincided with the results of numerical calculations for immobile atoms. The need for renormalization is due to the fact that for not very large optical thicknesses, it is necessary to substitute in the formula (\ref{5}) not $b_0$, but a slightly larger value. The difference is because of the extrapolation length for the boundary conditions of the radiative diffusion equation \cite{25}.
\begin{figure}
     \centering
     \includegraphics[width=0.5\textwidth]{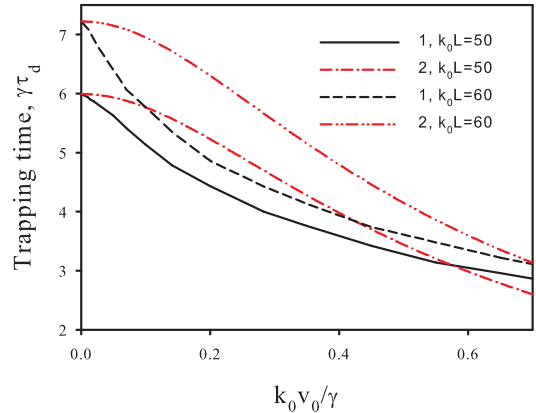}
    \caption{Dependence of $\tau_d$ on $k_0v_0$ (temperature) for two sizes of the atomic ensemble $k_0L=50$ and $k_0L=60$ respectively. The solid and dotted black lines  give the results of the numerical calculation. Red dash-dotted lines  are  drown on the base of Eq. (\ref{5}). }
\end{figure}

Figure 4 demonstrates a noticeable discrepancy between the results of the two calculations, which increases with an increase in the size of the ensemble. With heating, the discrepancy decreases, which, however, does not indicate a better applicability of the relation (\ref{5}). On the contrary, with heating, the optical thickness decreases and the diffusion approximation ceases to work. It formally predicts that $\tau_d$ tends to zero, while for optically thin media $\tau_d$ tends to the natural lifetime of atoms.

Our analysis shows that the detected discrepancy can be explained by the photon frequency drift during multiple scattering inside the cloud \cite{26, 27}. In the multiple scattering regime, a photon acquires a random frequency shift of order $k_0v_0$ at each scattering and its frequency performs a random walk in the frequency space. This frequency drift leads to the appearance of nonresonant photons, which have a large mean free path and, consequently, a shorter lifetime in the ensemble.

We analyzed the role of frequency diffusion by calculating the shape of the secondary radiation spectrum. The broadening of the spectrum of the secondary radiation was determined by a short-term Fourier transform \cite{28} with a rectangular window of duration $\gamma\Delta t = 30$. The center of the window was at times $\gamma t = 20$ after the end of the excitation pulse. The calculation results are shown in Fig. 5.
\begin{figure}
     \centering
     \includegraphics[width=0.5\textwidth]{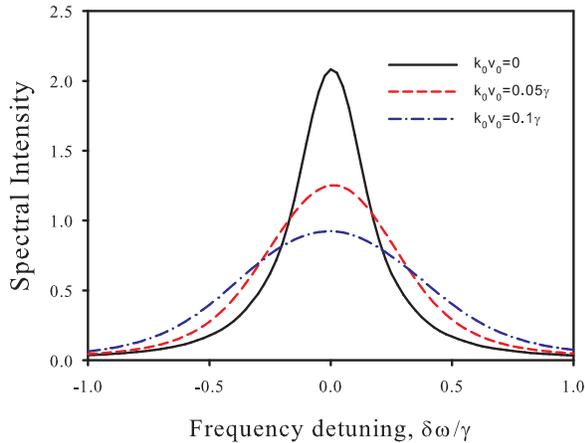}
\caption{Broadening of the fluorescence emission spectrum upon heating. The ensemble size is $k_0L=50$. The spectrum was calculated using the short-term Fourier transform. The center of the window corresponds to the time $\gamma t = 20$ }
\end{figure}

After the end of the excitation, the atoms begin to radiate at their own frequency. At times where the main mechanism is diffusion radiation trapping, there is a noticeable broadening of the spectrum due to multiple scattering. An increase in the size of the cloud and an increase in temperature enhance the effect of frequency drift, which explains the observed acceleration of fluorescence.

\subsection{Influence of motion on subradiation of dimers}
The role of motion manifests itself most unexpectedly at the stage of subradiation of diatomic clusters. As can be seen from Fig. 1 for all considered temperatures the motion reduces the duration of the trapping stage, and also, at certain time intervals, leads not to a weakening, but to an increase in the subradiation effect.

The influence of dimers begins to dominate  when the diffusion stage is completed. For the conditions for which Fig. 1 is drawn, this is the case for comparatively long times. The relative role of clusters can be enhanced if the diffusion effect is weakened. This can be done by reducing the optical thickness, since the influence of dimers do not need to be revealed against the background of diffusion trapping.

This is well demonstrated be the Fig. 6, which shows the dependence $\tau=\tau(t)$ for a fixed temperature corresponding to $k_0v_0=0.025\gamma$ and a fixed atomic density $nk_0^{-3}=0.005$, but for ensembles of different sizes having small optical thickness.
\begin{figure}
     \centering
     \includegraphics[width=0.5\textwidth]{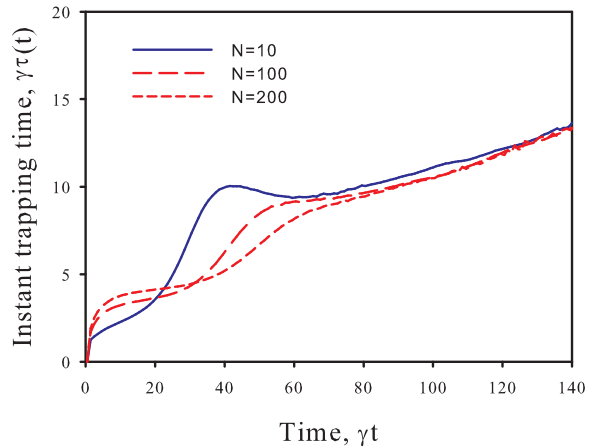}
\caption{Dynamics of instantaneous fluorescence delay time for various number of atoms. $k_0v_0=0.2\gamma$; $nk_0^{-3}=0.005$.  }
\end{figure}

Figure 6 shows another effect that appears when the motion is taken into account. For small systems, a nonmonotonic time dependence of the decay rate of the total fluorescence intensity is observed. At large times the curves $\tau=\tau(t)$ for ensembles of different sizes go to the same asymptote. This is due to the fact that the characteristic lifetime of long-lived excited states of atomic clusters depends on the average distance between atoms in them and does not depend on the size of the ensemble itself.

The main features of the influence of motion on cluster subradiation can be understood if we consider the temporal evolution of the excited state of a specific pair of atoms with a change in the distance between them. It is known that a system of two two-level atoms has six one-fold excited collective states. Two pairs of states are degenerate. The frequency shifts $\Delta_c$ and the width $\Gamma_c$ of the four distinct states of the stationary dimer can be found as follows
\begin{eqnarray}
\label{8}
\frac{\Delta_c}{\gamma}=\frac{3\epsilon}{4}\left( q\,\left( \frac{\cos(kr)}{(kr)^3}+\frac{\sin(kr)}{(kr)^2}\right) -\frac{p\cos(kr)}{kr}
\right),\nonumber\\ \\
\frac{\Gamma_c}{\gamma}=1-\frac{3\epsilon}{2}\left( q\,\left( \frac{\sin(kr)}{(kr)^3}-\frac{\cos(kr)}{(kr)^2}\right) -\frac{p\sin(kr)}{kr}  \right),
\nonumber
\end{eqnarray}
where $\epsilon=\pm1;\,p_0=0;\,q_0=-2;\, p_{\pm 1}=1;\, q_{\pm 1}=1$.

Let us consider how the total intensity as well as the population of the excited state of a diatomic quasimolecule changes with time, if atoms move and dimer is excited when interatomic distance is equal to given $r_0$.
\begin{figure}
     \centering
     \includegraphics[width=0.5\textwidth]{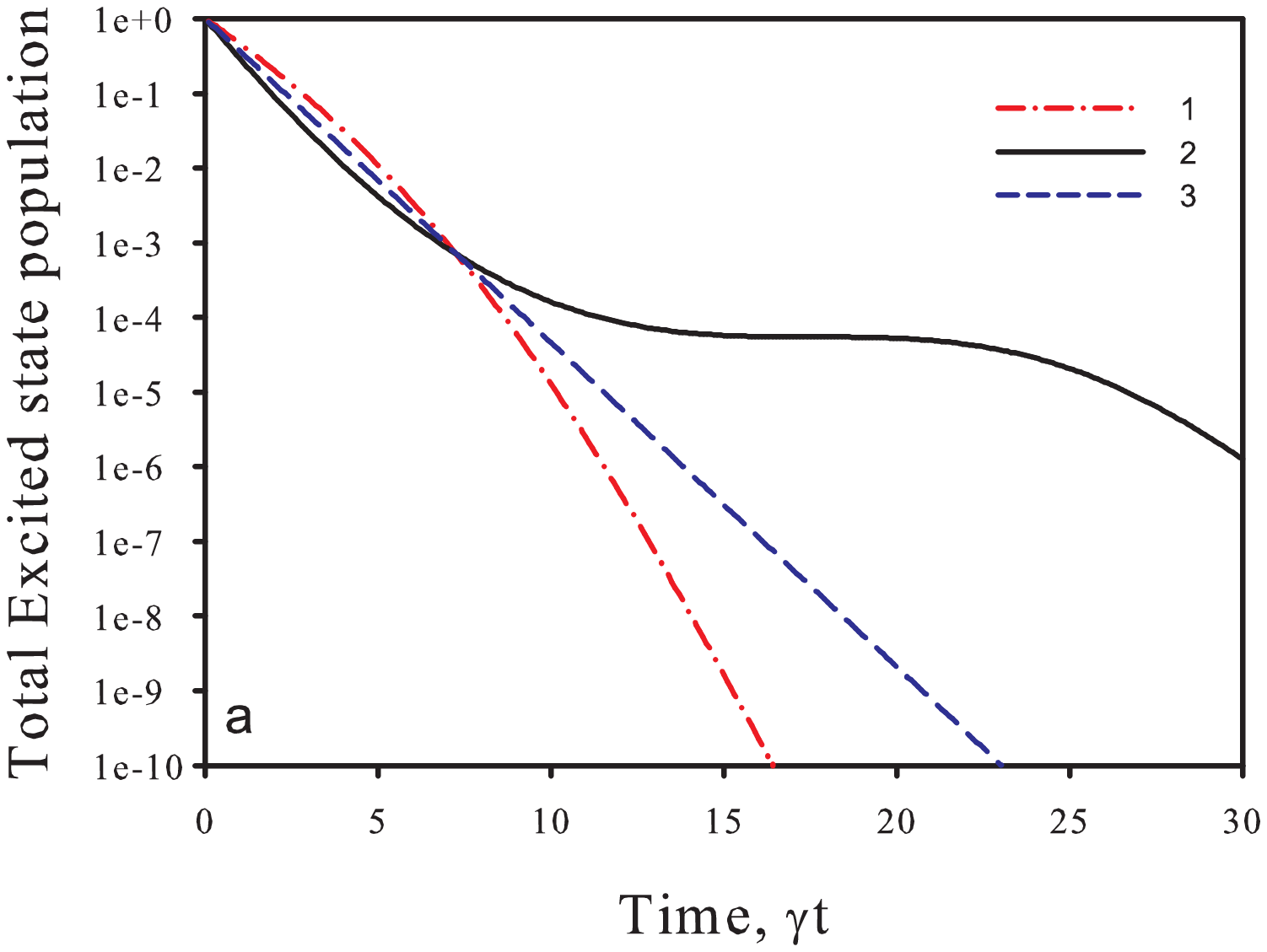}
          \includegraphics[width=0.5\textwidth]{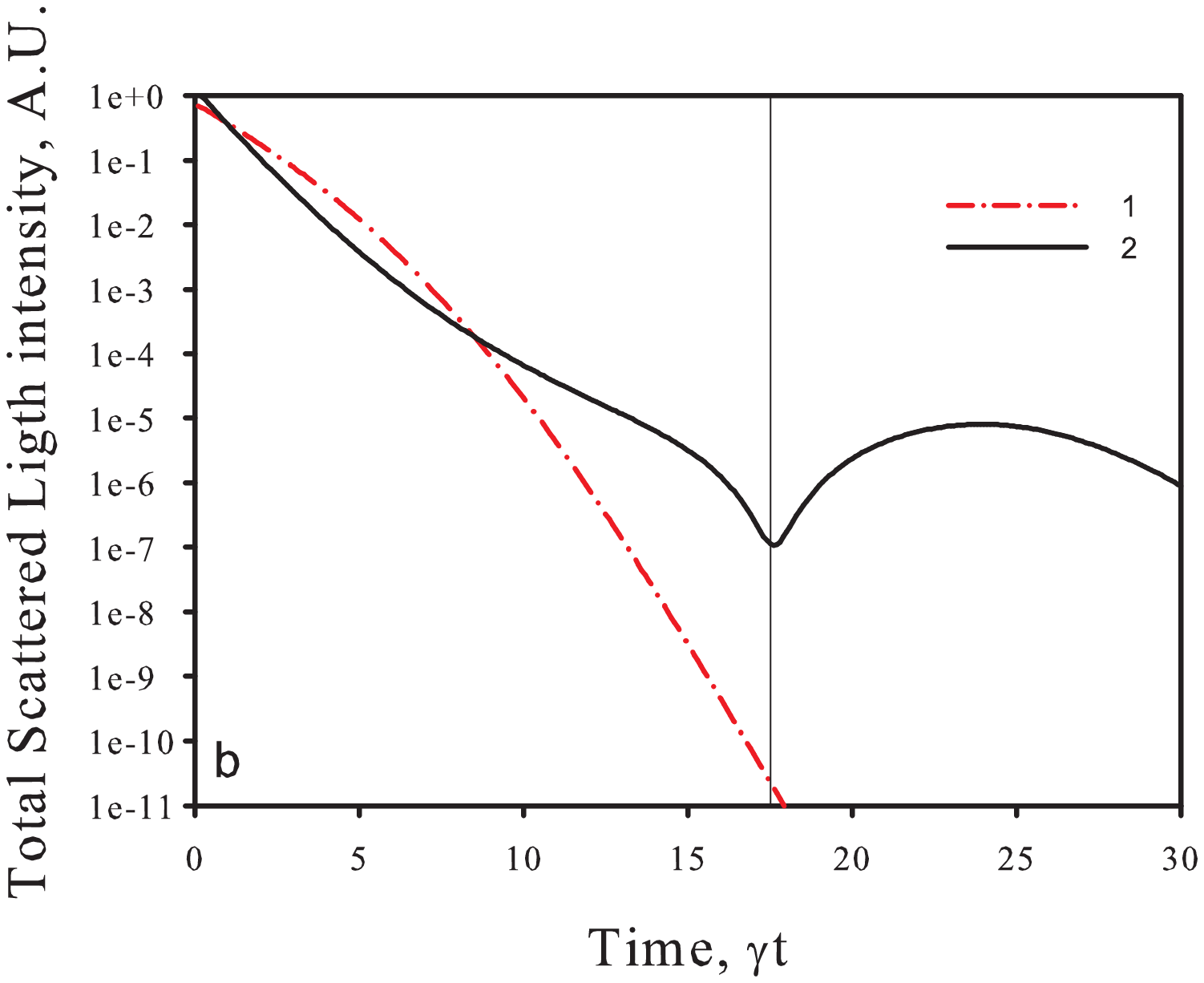}
\caption{a) Dynamics of the population of the excited state of a diatomic quasi-molecule with a change in the distance between atoms. b) Time dependence of total radiation intensity. Curve 1 corresponds to the initial excitation of the longest-lived state at $r_0=3.5k^{-1}_0$, curve 2 corresponds to the shortest-lived state. Curve 3 depicts  the decay with a rate of $\gamma$. Relative velocity of atoms is $k_0v=0.2\gamma$. Distance of closest approach is $r_m=0.1k^{-1}_0$. The vertical line corresponds to the moment of closest approach of the atoms. }
\end{figure}

Figures 7 shows the evolution of the considered system for two cases. Curves 1 and 2 correspond to the excitation of the longest-lived state and the shortest-lived one at $r_0$ correspondingly. For comparison curve 3 in Fig. 7a depicts  the decay of noninteracting atoms at a rate of $\gamma$. The curves are calculated for $r_0=3.5k^{-1}_0$. Distance of closest approach is $r_m=0.1k^{-1}_0$, relative velocity of atoms is $k_0v=0.2\gamma$.

Note that for the chosen conditions, the initially short-lived state (curve 2) becomes subradiant upon approach. At small interatomic distances the population of the excited state practically does not change. The radiation intensity decreases significantly. This manifests itself as a dip in curve 2 in Fig. 7b. After passing the point of closest approach, the radiation intensity increases. For the initial subradiante state, the picture is reversed. It decays very quickly when the atoms approach each other.

For motionless atoms each eigenstate of a quasimolecule decays independently of the others. Therefore, when any one of them is excited, other states are not populated during the further evolution of the system. The sub-radiant state remains sub-radiative. This is not the case for moving atoms. The relation (\ref{8}) describes the possibility of a subradiant state becoming superradiant even in the absence of transitions between different collective states. The decay rate of this state, i.e. of a state with given $\epsilon$, $p$, and $q$ varies nonmonotonically with $r$ and can be either greater or less than $\gamma$. This means that a change in the fluorescence rate of a cluster can be observed even in the absence of nonadiabatic transitions between its different states.

When atoms move, transitions between different collective states are also possible, which have an additional effect on the radiation dynamics. Such transitions are shown in Fig. 8. Shown here are the relative populations of four distinct states for different geometries of a diatomic quasi-molecule. It is assumed that one of the atoms is immobile. It is located at the origin of coordinates. The second atom moves parallel to the z axis. At the initial moment of time, he was at the point $k_0z_0=-3$, $k_0x_0=1$ $k_0y_0=0$. At this moment, the system is excited to the state which has the shift and width given by the formula (\ref{8}) with $\epsilon=1$, $p=0$ and $q=2$. Atom speed is equal to $k_0v=0.05\gamma$.
\begin{figure}
     \centering
     \includegraphics[width=0.5\textwidth]{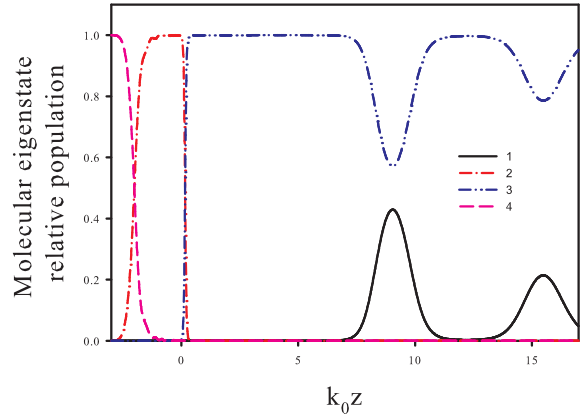}
\caption{Dynamics of the relative population of various collective states of a diatomic quasi-molecule with a change in the distance between atoms. One of the atoms is stationary, the second one moves with the speed $k_0v=0.05\gamma$ parallel to the $z$ axis. 1 corresponds to $\epsilon=1$, $p=1$ and $q=1$, 2: $\epsilon=-1$, $p=1$ and $q=1$, 3: $\epsilon=- 1$, $p=0$ and $q=2$, 4:$\epsilon=1$, $p=0$ and $q=2$. }
\end{figure}

The transitions between different eigenstates of a diatomic quasimolecule are clearly visible. Note that for the parameters under consideration, after the scattering of atoms, the most populated state is the longest-lived one with $\epsilon=-1$, $p=0$, and $q=2$. We checked that the last result is preserved regardless of which state was excited before the interatomic approach.

In a real multiatomic cloud, laser radiation excites not one, but a superposition of all possible states. And the nature of subradiation is determines by those of them that are subradiative at small interatomic distances. The others states decay rapidly and their population turns out to be low, which is manifested in the fluorescence of the ensemble as a whole.

\section{Conclusion}

In the present work, we study the effect of atomic motion on the dynamics of fluorescence of dilute atomic ensembles excited by resonant pulsed radiation. This effect is analyzed for three main stages of fluorescence evolution: at the stage of superradiance, the stage of diffuse trapping of radiation, and at the stage when subradiance is determined by the emission of atomic clusters randomly formed in the considered disordered atomic medium. It is shown that already for ensembles cooled to sub-Dplerian temperatures, motion can significantly affect the nature of the considered collective effects.

It is found that, in addition to an increase in the subradiation velocity into the coherent forward scattering cone, heating leads to the appearance of a nonmonotonic dependence of the radiation velocity. At certain time intervals, the decay of fluorescence in this direction can be replaced by its increase.
At the trapping stage, the main factor affecting the fluorescence rate is the diffusion of the secondary radiation frequency as a result of multiple scattering of light in an optically dense medium. We studied the fluorescence spectrum and revealed its significant broadening upon heating of the ensemble.
The most interesting results were found for subradiation of diatomic quasimolecules. In the temperature range corresponding to the MOT, the subradiation effect is enhanced for moving atoms. This effect is explained by the action of two factors. Firstly, by a change in the rate of decay of each of the eigenstates of a quasimolecule with a change in the distance between atoms, and, secondly, by possible nonadiabatic transitions between different sub- and superradiant states due to the motion of atoms.

\section{Acknowledgments}
The research was supported by a grant from the Foundation for the Development of Theoretical Physics and Mathematics "BASIS". The results of the work were obtained using the computing resources of the supercomputer center of Peter the Great St. Petersburg Polytechnic University (http://www.spbstu.ru).

\end{document}